\documentclass[a4paper,11pt]{article}

\usepackage{amsmath,amssymb}
\usepackage{tabularx}
\usepackage{epsfig}
\usepackage{graphicx}
\usepackage{fancyhdr}
\usepackage{cite}
\usepackage[left=3cm,top=2.5cm,right=3cm]{geometry}

\begin{document}
\begin{titlepage}
\begin{flushright} IFIC/11-10 \\  FTUV/11-0223 \end{flushright}
\vskip 2cm
\begin{center}
{\Large\bf The physics of a new gauge boson in a Stueckelberg \\[0.2cm] extension of the two-Higgs-doublet model}
\vskip 1cm
{\bf Grigoris Panotopoulos and Paula Tuz\'on}
\\[0.5cm]
Departament de F\'{i}sica Te\`{o}rica, IFIC, Universitat de Val\`{e}ncia - CSIC\\ Apt. Correus 22085, E-46071 Val\`{e}ncia, Spain
\\[0.5cm]
\end{center}
\vskip 2cm

\begin{abstract}
String theory constructions using D-brane
physics offer a framework where ingredients like extra abelian factors in the gauge group, more
than one Higgs doublet and a generalized Green-Schwarz mechanism appear at the same time.
Motivated by works towards the direction of obtaining the Standard Model in orientifold
constructions, we study in the present work a Stueckelberg extension of the two-Higgs-doublet model.
The distinctive features of our model are i) a sharp decay width for the heavy gauge boson, and ii)
a charged Higgs boson having two main decay channels at tree level with equal branching ratios.
\end{abstract}

\end{titlepage}

\section{Introduction}

The Standard Model
of Particle Physics (SM) (for reviews see e.g.~\cite{Langacker:1995hi,Pich:2007vu})
has been
extremely successful in
describing all low energy phenomena, being in excellent agreement with
a vast amount of experimental data. The only missing part of the SM today is the
Higgs boson that gives masses to fermions and to $W^{\pm}$ and $Z$ bosons. The Stueckelberg
mechanism~\cite{Stueckelberg:1938zz} gives mass to abelian vector bosons without
breaking gauge invariance on the Lagrangian, and thus provides an alternative
to the Higgs mechanism~\cite{Higgs:1964ia,Higgs:1964pj,Higgs:1966ev,Guralnik:1964eu,Englert:1964et} to achieve gauge symmetry breaking without spoiling
renormalizability. A Stueckelberg extension of the SM was studied in~\cite{Kors:2004dx,Feldman:2006ce} (see also~\cite{Feldman:2007wj}
for a generalization with a kinetic mixing),
where
the neutral electroweak gauge bosons aquire a mass via both the Higgs and the
Stueckelberg mechanism.

Most of the well motivated extensions of the SM, which have been developed to
address its open issues, involve an extra $U(1)$ in the gauge group. A new heavy
gauge boson, $Z^\prime$, is predicted which would have profound implications for
particle physics and cosmology. Such gauge bosons occur naturally in $SO(10)$ grand
unified models, extra dimensional models with a hidden sector brane, and string theoretic
models with intersecting branes. For a nice recent review see e.g.~\cite{Langacker:2008yv}.
In general, in models with an additional abelian factor in the gauge group the fermions
are charged under the extra $U(1)$, and furthermore there is a mixing between the SM
boson $Z$ and the new gauge boson $Z^\prime$. A $Z^\prime$ boson that mixes with the
SM $Z$ boson distorts its properties, such as couplings to
fermions and mass relative to electroweak inputs. One thus has to worry
about the cancellation of all anomalies, and remain in agreement with the LEP
and SLC data~\cite{Carena:2004xs}.

Another famous minimal extension of the SM consists in the addition of one scalar doublet to the theory \cite{Lee:1973iz}. This idea has been particularly successful for its simplicity and the rich phenomenology that generates, being able to introduce new dynamical possibilities, like different sources of CP violation or dark matter candidates, helps to solve some of the SM problems.
In the most general version of the two-Higgs-doublet model (2HDM), the fermionic couplings of the neutral scalars are not diagonal in flavour, which generates dangerous flavour-changing neutral current (FCNC) phenomena. Since these are tightly constrained by the experimental data, it is necessary to implement ad-hoc dynamical restrictions to guarantee their absence at the required level. For that aim, several versions of the 2HDM based in different ideas have been developed: One possibility is to assume particular Yukawa textures generated by some flavour symmetry, which force the non-diagonal Yukawa couplings to be very small \cite{Cheng:1987rs,Atwood:1996vj,DiazCruz:2004pj,DiazCruz:2009ek}. Another elegant solution comes from considering the alignment of both scalar doublets in flavour space \cite{Pich:2009sp}, this implies that the only flavour-changing source is the Cabibbo-Kobayashi-Maskawa (CKM) quark mixing matrix in the charged sector and all neutral couplings are diagonal; moreover, this approach provides interesting new sources of $CP$ violation. Particular cases of this general idea are the models defined by the implementation of a discrete $\mathcal Z_2$ symmetry, preserving $CP$ and making only one scalar doublet to couple to a given right-handed fermion sector \cite{Glashow:1976nt}. Depending on the way this symmetry is implemented, different kind of $\mathcal Z_2$ models are generated \cite{Haber:1978jt,Hall:1981bc,Donoghue:1978cj,Barger:1989fj,Savage:1991qh,Grossman:1994jb,Akeroyd:1998ui,Akeroyd:1996di,Akeroyd:1994ga,Aoki:2009ha,Ma:2008uza,Ma:2006km, Barbieri:2006dq,LopezHonorez:2006gr}.
Another possibility is to study the general version of the model in the decoupling limit, where the masses of the new Higgs bosons are very heavy and suppress \emph{naturally} the FCNC couplings.
In this sense, and as will be seen in the following, this work opens another way of having a reliable 2HDM in the presence of an extra $U(1)$ gauge symmetry. The only possible 2HDM within this framework has not FCNC terms and the scalar potential results $CP$ invariant.

Many attempts have been made in order to embed the
SM in open string theory, with some success~\cite{Antoniadis:2000ena,Bailin:2000kd,Ibanez:2001nd,Kokorelis:2002ip,Kokorelis:2002zz,Cvetic:2002qa,Bailin:2002gg,Kokorelis:2002xm,Bailin:2002ti,Antoniadis:2002qm}. They consider
the SM particles as open string states attached on different stacks of D-branes. $N$
coincident D-branes typically generate a unitary group $U(N) \sim SU(N) \times U(1)$.
Therefore, every stack
of branes supplies the model with an extra abelian factor in the gauge group.
Such $U(1)$ fields have generically four-dimensional anomalies~\cite{Anastasopoulos:2003aj,Anastasopoulos:2004ga}.
These anomalies are cancelled via the Green-Schwarz mechanism~\cite{Green:1984sg,Green:1984qs,Sagnotti:1992qw,Ibanez:1998qp} where
a scalar axionic
field is responsible for the anomaly cancellation.
This mechanism gives a mass to the anomalous $U(1)$ fields and breaks the associated
gauge symmetry. If the string scale is around a few TeV, observation of such anomalous
$U(1)$ gauge bosons becomes a realistic possibility~\cite{Kiritsis:2002aj,Ghilencea:2002da,Ghilencea:2002by}. The structure
of the Minimal Low Scale Orientifold Model has been presented in detail
in~\cite{Coriano':2005js}.

This class of models is characterized by i) the existence of two Higgs doublets necessary
to give masses to all fermions, and ii) the massive
gauge bosons acquire their mass from two sources, namely the usual Higgs mechanism, as well
as the stringy mechanism related to the generalized Green-Schwarz mechanism, which is
very similar to the Stueckelberg mechanism.
In the light of these developments, it becomes clear that it is natural to study the 2HDM
with additional $U(1)$s and the Stueckelberg mechanism together with the Higgs mechanism.
In the present work we wish to study the phenomenology of a simple four-dimensional, non-GUT,
non-supersymmetric model with an additional Higgs doublet, and just one extra $U(1)$ factor
in the gauge group for simplicity. In a similar spirit, albeit in a different set-up,
possible signatures at colliders of new invisible
physics and Stueckelberg axions have been analyzed in~\cite{Coriano:2007xg,Kumar:2007zza,Armillis:2008vp,Antoniadis:2009ze,Coriano:2009zh}.

Our work is summarized as follows. In the next section we present the model,
in section 3 we discuss the physics of the heavy gauge boson, and finally we
conclude in section 4.

\section{The model}

Here we shall present the ingredients of the model, the electroweak symmetry breaking, and
the mass spectrum at tree level, while the relevant interaction vertices will be given
in the next section.

The gauge group of the model is the SM gauge group times an extra abelian
factor $U(1)_X$, with a coupling constant $g_X$ and a gauge boson $C_\mu$
associated with it. We have three generations of quarks and leptons
with the usual quantum numbers under the SM gauge group, and they are
assumed to be neutral under the extra $U(1)$. This is a simple choice that ensures
that there are no anomalies in the model. We consider the presence
of two Higgs
doublets, $H_1$ and $H_2$,
with the same quantum numbers under the SM gauge group, the only difference
being is that $H_1$ is assumed to be neutral under
$U(1)_X$, while $H_2$ is charged
under the
additional abelian factor with charge $Y_X= \pm 1$\footnote{In intersecting
brane models one naturally obtains
two Higgs doublets, with three possibilities arising regarding their charges under the
$U(1)$ factors: i) both are neutral, ii) one is neutral and one charged, and iii)
both are charged with opposite charges~\cite{Kiritsis:2002aj,Antoniadis:2000ena,Antoniadis:2002qm}. The
first possibility is not interesting, while in
the third one no Yukawa couplings are allowed.}.
As a consequence no Yukawa terms including $H_2$
are allowed by the symmetry, and therefore the FCNC problem is avoided.
The gauge interactions
are thus completely specified, and the Yukawa couplings are the same as in the SM. The most general
Higgs potential which is renormalizable and compatible with the symmetries in this framework is the following:
\begin{eqnarray}
V & = & \mu_{1}^{2}H_{1}^{\dagger}H_{1}+\mu_{2}^{2}H_{2}^{\dagger}H_{2}+\frac{1}{2}\lambda_{1}(H_{1}^{\dagger}H_{1})^{2}+\frac{1}{2}\lambda_{2}(H_{2}^{\dagger}H_{2})^{2}+\lambda_{3}(H_{1}^{\dagger}H_{1})(H_{2}^{\dagger}H_{2})\nonumber \\
 & + & \lambda_{4}(H_{1}^{\dagger}H_{2})(H_{2}^{\dagger}H_{1}) \; ,
\end{eqnarray}
where $\mu_{1,2}$ and $\lambda_{1-4}$ are real parameters. This corresponds to an \emph{inert} approach, with the potential similar to the one generated by imposing a $\mathcal Z_2$ discrete symmetry in the \emph{Higgs basis} of a general two-Higgs-doublet model with the SM gauge group \cite{Ma:2008uza,Ma:2006km, Barbieri:2006dq,LopezHonorez:2006gr}, but in this case, resulting from a gauge symmetry which additionally forbids the $\lambda_5$ term. From the three possible solutions given by the minimization conditions of the potential~\cite{Deshpande:1977rw}, we choose for analogy the one where the second Higgs doublet vacuum expectation value (VEV) is zero, $<0|H_2|0>=0$, and only the first Higgs doublet, $H_1$, acquires a VEV, $v$.
As in the $\mathcal Z_2$-inert model, the nonexistence of a VEV for $H_2$ ensures the
absence of mixing between the components
of $H_1$ and those of $H_2$. Hence, $H_1$ closely corresponds to the ordinary SM Higgs doublet,
and the fields belonging to $H_2$ are inert in the sense that they do not couple directly to
fermions, but they have gauge interactions and self-interactions.

Finally, the Stueckelberg contribution is~\cite{Kors:2004dx}
\begin{eqnarray}
{\cal L}_{\rm St} = -\frac{1}{4} C_{\mu\nu}C^{\mu\nu}
- \frac{1}{2} (\partial_{\mu}\sigma  + M_1 C_{\mu} + M_2 B_{\mu})^2 \; ,
\end{eqnarray}
where $C_\mu$ is the gauge boson associated with the $U(1)_X$, $C_{\mu \nu}$ is the
corresponding field strength, $\sigma$ is the scalar axionic field which is assumed to couple
both to $B_\mu$ and $C_\mu$, and $M_1$ and $M_2$ are two mass scales which serve as two extra
parameters of the model. \\

After giving masses to the gauge bosons, the doublets are of the form~\cite{Barbieri:2006dq}
\begin{eqnarray}
H_1 = \left [  \begin{array}{c} 0 \\  \frac{1}{\sqrt{2}} (v + h)  \end{array}  \right] \;  \qquad H_2 = \left [  \begin{array}{c} H^+ \\  \frac{1}{\sqrt{2}} ( H+ iA)  \end{array}  \right] \; ,
\end{eqnarray}
where $H_1$ has one physical degree of freedom left: the
neutral scalar field $h$.
Since $h$ closely resembles the Higgs particle of the SM
it will be called here the SM
Higgs boson. In addition, $H_2$ includes the neutral $CP$-even
$H$, the neutral $CP$-odd $A$ (with defined $CP$ parities because the parameters of the scalar potential are real), and two charged $H^{\pm}$
inert scalars. The masses of the particles are
(at tree level) given by
\begin{eqnarray}
M^{2}_h & = & \lambda_{1}v^{2} \label{eq:mass-sm-higgs} \nonumber \\
M^{2}_{H^{\pm}} & = & \mu_{2}^{2}+\frac{1}{2} \lambda_{3}v^{2} \label{eq:mass-HP} \nonumber \\
M^{2}_H& = & \mu_{2}^{2}+\frac{1}{2} (\lambda_{3}+\lambda_{4})v^{2} \label{eq:mass-H0} \nonumber \\
M^{2}_A & = & \mu_{2}^{2}+\frac{1}{2} (\lambda_{3}+\lambda_{4})v^{2} \label{eq:mass-A0} \; .
\end{eqnarray}
Notice that in our model the neutral inert Higgs bosons are exactly degenerate in mass, and this should
be true also at loop level, since $A$ and $H$ couple to the same fields with the same coupling constants.
If $H$ and $A$ are exactly degenerate in mass, there is not a good dark matter candidate because of direct
detection limits~\cite{LopezHonorez:2006gr}. In direct detection searches the dark matter particle scatters
off a nucleous of the
material in the detector. What is seen is the recoil of the nucleous, while the dark matter particle is
not observed. Analysing the data an upper bound on the nucleon/dark matter particle cross section for
a given dark matter particle mass is obtained~\cite{CDMS:2011gh}. If the masses of $H,A$ are different, the
lightest of the two Higgs bosons, say $H$, is supposed to play the role of dark matter in the universe, and
the scattering off a nucleous takes place via an exchange of the SM Higgs boson (see the first Feynman
diagram in the Figure~7 below). If, however, the two inert
Higgs bosons are degenerate in mass, the scattering of $H$ off a nucleous can take place also via a $Z$ boson
exchange (see the second Feynman diagram in the Figure~7 below). In this case there is an unsuppressed
coupling with the $Z$ boson, and the elastic scattering $H q \rightarrow A q$ through
a $Z$ boson exchange has a cross section orders of magnitude larger than the allowed
ones~\cite{LopezHonorez:2006gr}.
\begin{figure}[tbh!]
\begin{center}
\begin{tabular}{cc}
\includegraphics[width=7cm]{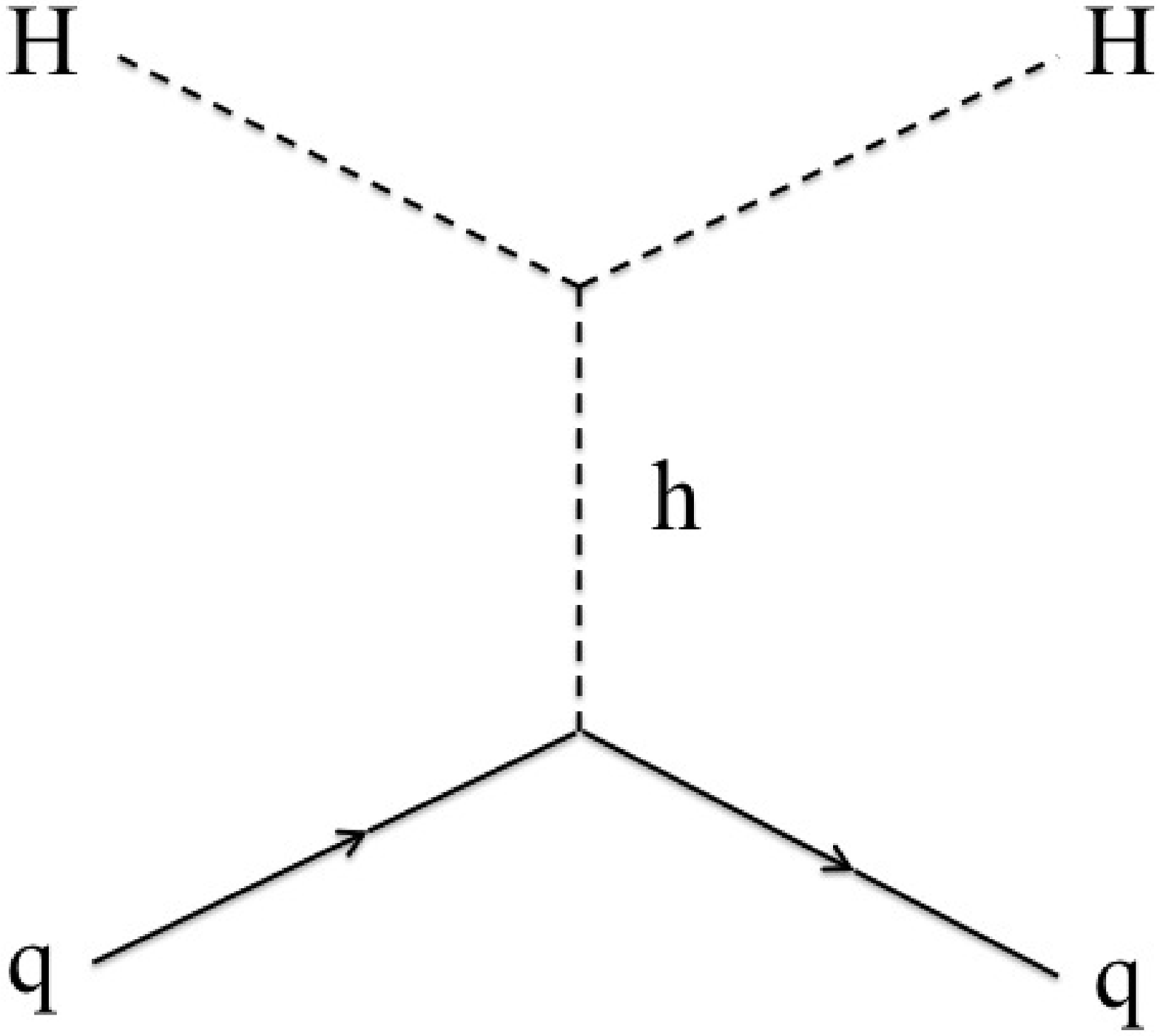} & \includegraphics[width=7cm]{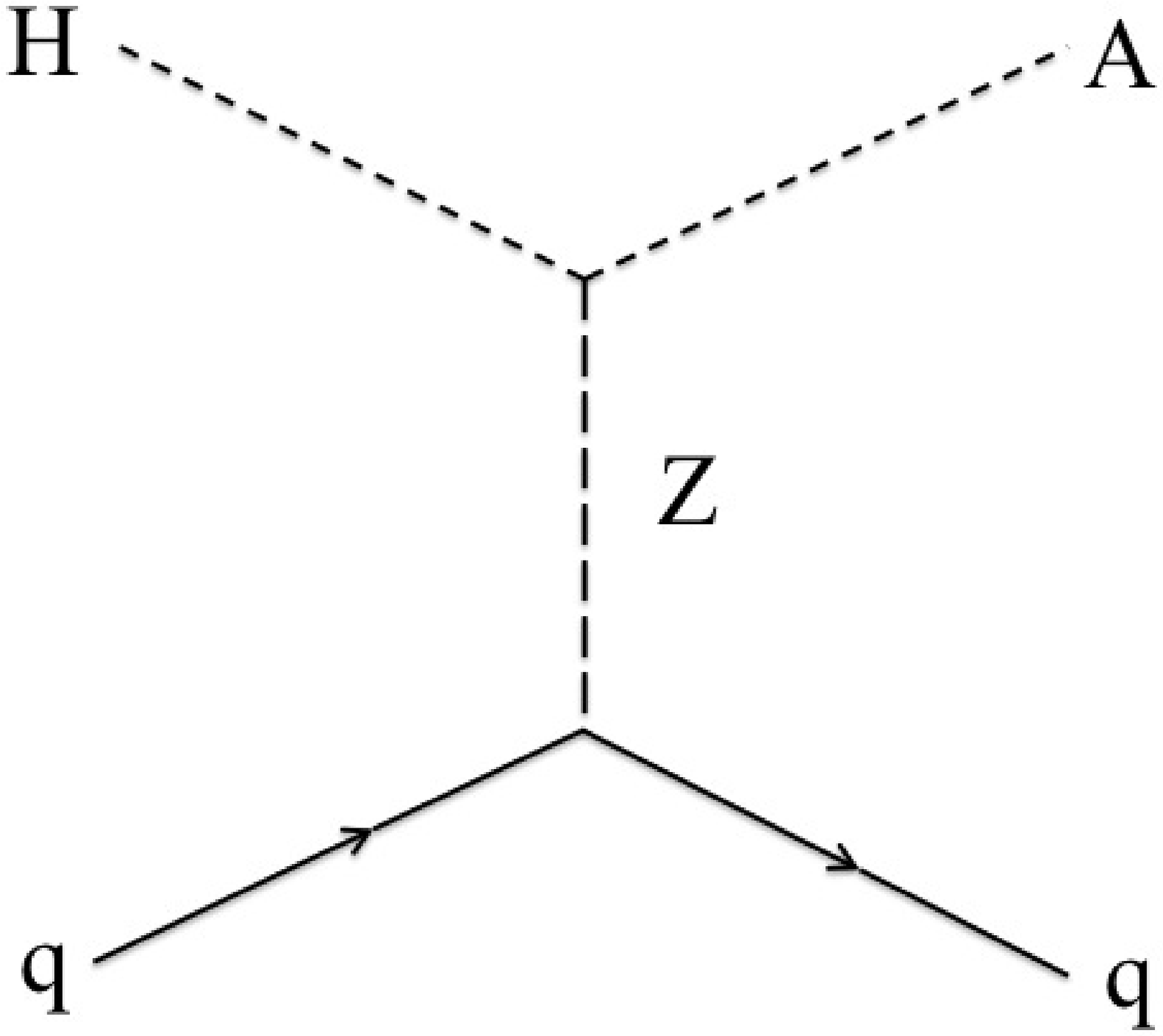}
\end{tabular}
\caption{Feynman diagrams corresponding to the elastic scatterings $H \; q \to H \; q$ and $H \; q \to A \; q$ respectively.\label{feyn}}
\end{center}
\end{figure}
The components of the inert
scalar doublet interact with $h$ and among themselves as follows:
\begin{eqnarray}
V_{int} & = & \frac{1}{2}\lambda_{2}\left[H^{+}H^{-}+\frac{1}{2}H^{2}+\frac{1}{2}A^{2}\right]^{2} \nonumber \\
& + & \lambda_{3}\left(vh+\frac{1}{2}h^{2}\right)\left[H^{+}H^{-}+\frac{1}{2}H^{2}+\frac{1}{2}A^{2}\right]\nonumber \\
 & + & \frac{1}{2}\lambda_{4}\left(vh+\frac{1}{2}h^{2}\right)H^{2}+\frac{1}{2}\lambda_{4}\left(vh+\frac{1}{2}h^{2}\right)A^{2} \; .
\end{eqnarray}
With the Stueckelberg extension, and
after the standard spontaneous electroweak symmetry breaking the
mass terms in the neutral vector boson sector take the form
$-\frac12 V_{a\mu} M^2_{ab} V^\mu_b$, using $(V_\mu^{\rm
T})_a = ( C_{\mu}, B_{\mu}, W_{\mu}^3)_a$, with mass matrix~\cite{Kors:2004dx}
\begin{eqnarray}
M^2_{ab} =
\left[\begin{matrix} M_1^2 & M_1M_2 & 0 \\
M_1M_2 & M_2^2 + \frac{1}{4} g_Y^2 v^2 & - \frac{1}{4} g_Y g_2 v^2 \\
0 & -\frac{1}{4} g_Y g_2 v^2 & \frac{1}{4} g_2^2 v^2 \end{matrix}\right] \; ,
\end{eqnarray}
where $v=2M_{\rm W}/g_2=(\sqrt 2 G_F)^{-\frac{1}{2}}=246$~GeV, $g_2$ and
$g_Y$ are the $SU(2)_L\times U(1)_Y$ gauge coupling constants,
$M_{\rm W}$ is the mass of the $W^{\pm}$ boson and $G_F$ is the Fermi
constant. From $\det(M_{ab}^2)=0$ it is easily seen that one eigenvalue
is zero, whose eigenvector
corresponds to the photon
$A_\mu^\gamma$. Among the remaining two eigenvalues $M_{\pm}^2$,
we identify the lighter mass eigenstate with mass $M_-$ as the
$Z$ boson, and the heavy eigenstate with mass $M_+$ as the
$Z'$ boson. Using an orthogonal transformation $O$ to diagonalize $M_{ab}^2$, ${O}^T  M_{ab}^2  {O}=M_{\rm D}^2$,
we go to the eigenstates basis $E_{\mu}^T=({\rm Z}_{\mu}',{\rm Z}_{\mu},A_{\mu}^\gamma)$, where $M^2_{\rm D}={\rm diag}(M_{{\rm Z}'}^2, M_{\rm Z}^2,0)$, $V_{\mu} = O E_{\mu}$ and $O$ is parametrized as
\begin{eqnarray}\label{rot}
{O}=
\left[\begin{matrix} \cos\psi \cos\phi -\sin\theta\sin\phi\sin\psi &
-\sin\psi \cos\phi -\sin\theta\sin\phi\cos\psi & -\cos\theta \sin\phi\cr
\cos\psi \sin\phi +\sin\theta\cos\phi\sin\psi &
-\sin\psi \sin\phi +\sin\theta\cos\phi\cos\psi & \cos\theta \cos\phi\cr
-\cos\theta\sin\psi & -\cos\theta\cos\psi & \sin\theta \end{matrix}\right] \; .
\end{eqnarray}
The mixing angles $\theta$, $\phi$ and $\psi$ are given by~\cite{Kors:2004dx}
\begin{eqnarray}
\tan \theta = \frac{g_Y}{g_2}\cos \phi\ , ~
\tan \phi = \frac{M_2}{M_1} \ , ~
\tan \psi =
\frac{\tan \theta \tan\phi M_{\rm W}^2}
     {\cos\theta(M^2_{{\rm Z}'}-(1+\tan^2\theta)M_{\rm W}^2)}\ \; ,
\end{eqnarray}
and we can again define the weak angle to be the same as in the SM,
$\tan\theta_w=g_Y/g_2$.
In this model, there are some extra free parameters apart from the SM ones,
which are: i) the mass parameters and couplings in the Higgs potential, and ii)
the coupling constant $g_X$ and the mass scales $M_1, M_2$ (or $M_2/M_1$
and $M_{Z^\prime}$) from the Stueckelberg
contribution. Usually, to be consistent with the LEP data on the $Z$ boson, the mixing between the $Z$ and $Z^\prime$
bosons has to be small, $|\epsilon| \leq 0.001 $~\cite{Carena:2004xs}, which is satisfied when either the $Z^\prime$
boson is heavy or the new coupling constant $g_X$ is very small. In the model discussed here, the couplings of $Z^\prime$ to the
fermions (see the relevant formulas in the next section) do not deviate significantly from the SM values
if we take a small ratio $M_2/M_1 \leq (0.05-0.06)$~\cite{Feldman:2006ce}. We have considered mainly the case in which
$M_2/M_1 = 0.03$, but later on we will also make a comment on what the effect of varying the ratio is.
Since $M_2/M_1$ is taken to be small, the new gauge boson is allowed to be relatively light, and if not very
heavy it is within future experimental reach. Therefore in the following we shall take the mass
of the heavy gauge boson to be in the range between $200$~GeV and $1-2$~TeV.

The interactions between the fermions, the SM Higgs and the charged $W^{\pm}$ bosons
are the same as in the SM, and the electromagnetic interactions of charged
particles have the usual form, where now the electric charge is given by
\begin{equation}
e=\frac{g_2 g_Y \cos(\phi)}{\sqrt{g_2^2+g_Y^2 \cos(\phi)^2}} \; .
\end{equation}
Furthermore, the interactions of the inert bosons with $W^{\pm}$ are the same
as in the inert 2HDM. However, due to the existence of $C_\mu$ and the new mixing
between the mass eigenstates and the
gauge eigenstates, the couplings to
the $Z$ boson are different, and there are also similar couplings to the $Z^\prime$.
In the model discussed here, the photon is a linear combination of $W_\mu^3$ ,$C_\mu$ ,$B_\mu$,
and we find the relation
\begin{equation}
Q=T_3+\frac{Y}{2}-\frac{g_X Y_X M_2}{2 g_Y M_1}
\end{equation}
which generalizes the usual SM formula $Q=T_3+Y/2$. For the particles that are neutral
under the extra $U(1)$ factor the third term vanishes and we recover the formula valid in
the SM, while for the inert Higgs bosons, assuming that $Y_X=\pm 1$, we find
\begin{equation}
Y=1 \pm \frac{g_X}{g_Y} \: \frac{M_2}{M_1}
\end{equation}
and therefore for a coupling constant $g_X$ similar to the SM coupling constants $g_2$, $g_Y$
or lower, the usual hypercharge for the inert Higgs bosons is slightly different than one.

\section{Heavy gauge boson searches}

The LHC is designed to collide protons with a center-of-mass energy
14 TeV. Since the center-of-mass energy of proton-proton
collisions at LHC is 14 TeV, the particle cascades
coming from the collisions might contain $Z^\prime$ if its mass
is of the order of 1 TeV. Therefore a heavy gauge boson can be discovered at LHC, and in fact
new gauge bosons are perhaps the next best motivated new physics, after the Higgs
and supersymmetric particles, to be searched for at future experiments.
The mass, total decay width as well as branching ratios for various decay modes are some
of the properties of  $Z^\prime$ that should be accurately measurable, and could be used to distinquish
between various models at colliders. Thus, in this section we discuss the phenomenology of the model as
far as the physics of the new gauge boson is concerned.
In the following we shall be interested
in two-body decays, $M \rightarrow m_1 \; m_2$, of a heavy particle with mass $M$ into two
lighter particles with masses $m_1, m_2$, provided of course that in the Lagrangian there is the
corresponding three-point vertex, and that the decay is kinematically
allowed, namely that $M > m_1+m_2$. The general formula for the decay width
is given by
\begin{equation}\label{genrate}
\Gamma(M\rightarrow m_1\; m_2)=\frac{\lambda^{1/2}(M^2,m_1^2,m_2^2)}{16 \pi M^3} |\mathcal{M}_{fi}|^2 \; ,
\end{equation}
where $\mathcal M_{fi}$ is the transition amplitude from the initial to final state and the function $\lambda(a,b,c) \equiv a^2+b^2+c^2-2ab-2ac-2bc$.
In particular, we are here interested in the decays of the heavy gauge boson $Z^\prime$
into fermions, $W^{\pm}$, $Z$ and Higgs bosons:
\begin{eqnarray}\label{channels}
Z^\prime  & \rightarrow & f \bar f \nonumber \\
Z^\prime  & \rightarrow & W^+ W^- \nonumber \\
Z^\prime  & \rightarrow & H^+ H^-\nonumber \\
Z^\prime  & \rightarrow & H A \nonumber \\
Z^\prime  & \rightarrow & h \;  Z \; .
\end{eqnarray}

\subsection{$Z'$ to fermions}

For the first decay channel in \eqref{channels}, the Lagrangian interaction between fermions and a massive neutral
gauge boson $V$ has the form
\begin{equation}
\mathcal L_{Vff}=-g_{Vff} \bar{f} \gamma^\mu (c_V+c_A \gamma^5)f V_\mu \; ,
\end{equation}
and the corresponding decay width is given by
\begin{equation}
\Gamma(V   \rightarrow  f \bar{f})=N_c \frac{g_{Vff}^2 M_V}{12 \pi} \left[ c_V^2+c_A^2 +2(c_V^2-2c_A^2) \frac{m_f^2}{M_V^2} \right] \sqrt{1-4\frac{m_f^2}{M_V^2}} \; ,
\end{equation}
which in the massless fermion limit ($M_V \gg m_f$) is simplified to
\begin{equation}
\Gamma(V  \rightarrow  f \bar{f})=N_c \frac{g_{Vff}^2 (c_V^2+c_A^2)M_V}{12 \pi} \; ,
\end{equation}
where the number of colors $N_c$ is one for leptons and three for quarks.
In the model discussed here $V$ corresponds to the $Z'$ boson and the coupling $g_{Z^\prime ff}$ is given by
\begin{equation}
g_{Z^\prime ff}=\frac{g_2}{4 \cos(\theta_w)} \; ,
\end{equation}
while $c_V$ and $c_A$ are computed to be
\begin{eqnarray}
c_V & = & 2T_3\cos(\theta_w)O_{31}+(Y_L+Y_R)\sin(\theta_w)O_{21} \nonumber \\
c_A & = & -2T_3\cos(\theta_w)O_{31}-(Y_L-Y_R)\sin(\theta_w)O_{21} \; ,
\end{eqnarray}
and can be also found in \cite{Cheung:2010az}. In Table
\ref{charges} we recall the quantum number of the fermions.
\begin{table}[h!]
\begin{center}
\begin{tabular}{|l|c|c|c|}
\hline
&$T_3$& $Y_L$ & $Y_R$ \\
\hline
Neutrinos &$\frac{1}{2}$&$-1$& $0$\\
Charged leptons &$-\frac{1}{2}$&$-1$& $-2$\\
Up quarks &$\frac{1}{2}$&$\frac{1}{3}$& $\frac{4}{3}$\\
Down quarks &$-\frac{1}{2}$&$\frac{1}{3}$& $-\frac{2}{3}$\\
\hline
\end{tabular}
\caption{\label{charges} Fermion quantum numbers.}
\end{center}
\end{table}

\subsection{$Z'$ to bosons}

For the decay channels of $Z^\prime$ into Higg bosons the Lagrangian interaction has the
usual structure as in~\cite{Barger:1987xw}. Departing from the neutral gauge eigenstates $V_{\mu} = \left\{ C_{\mu} ,B_{\mu}, W^3_{\mu} \right\}$ Lagrangians,
\begin{eqnarray}
i \mathcal L_{V H^+H^-} &=& \frac{g_Y}{2}  \left( H^+ \stackrel{\leftrightarrow} {\partial}_{\mu} H^-   \right) B^{\mu} + \frac{Y_{X}}{2} g_{X} \left( H^+ \stackrel{\leftrightarrow} {\partial}_{\mu} H^-   \right) C^{\mu}
+ \frac{g_{2}}{2}  \left( H^+ \stackrel{\leftrightarrow} {\partial}_{\mu} H^-   \right) W^{\mu}_3  \nonumber \\
\mathcal L_{V HA} &=&  \frac{ g_Y}{2} \left( H \stackrel{\leftrightarrow} {\partial}_{\mu} A   \right) B^{\mu} + \frac{Y_{X}}{2}g_{X}  \left( H \stackrel{\leftrightarrow} {\partial}_{\mu} A   \right) C^{\mu}
- \frac{g_{2}}{2}  \left( H \stackrel{\leftrightarrow} {\partial}_{\mu} A  \right) W^{\mu}_3\nonumber \\
\mathcal L_{V Vh} &=& \frac{1}{v} \; h \;  M_Z^2  \left( \cos(\theta_w) W_{\mu}^3 -\sin(\theta_w)B_{\mu} \right)  \left( \cos(\theta_w) W^{\mu}_3 -\sin(\theta_w)B^{\mu} \right) \; ,
\end{eqnarray}
one can derive the interaction between the new gauge boson $Z^\prime$ rotating the fields as in Eq. \eqref{rot}. Therefore, from the mass eigenstates basis $E_{\mu}=\left\{ Z_{\mu}^\prime ,Z_{\mu}, A^{\gamma}_{\mu} \right\}$, the contributions describing these interactions are,
\begin{eqnarray}
\mathcal L_{Z^\prime H^+H^-} &=&-i\; g_{Z^\prime H^+ H^-}  \;  Z'^{\mu} \left( H^+ \stackrel{\leftrightarrow} {\partial}_{\mu} H^-   \right) \nonumber \\
\mathcal L_{Z^\prime HA} &=&g_{Z^\prime H A} \;  Z'^{\mu} \left( H \stackrel{\leftrightarrow} {\partial}_{\mu} A   \right)  \nonumber \\
\mathcal L_{Z^\prime Zh} &=&g_{Z^\prime Z h}\;  Z'^{\mu} Z_{\mu} h \; ,
\end{eqnarray}
given the corresponding couplings,
\begin{eqnarray}
g_{Z^\prime H^+ H^-} & = & \frac{1}{2}(g_2 O_{31}+g_Y Y O_{21}+g_X Y_X O_{11})\nonumber \\
g_{Z^\prime H A} & = & \frac{1}{2}(-g_2 O_{31}+g_Y Y O_{21}+g_X Y_X O_{11})\nonumber \\
g_{Z^\prime Z h} & = & \frac{M_Z^2}{v} ( 2\cos^2(\theta_w)O_{31}O_{32}+2\sin^2(\theta_w)O_{21}O_{22} \nonumber \\
& - & \sin(2\theta_w) (O_{31} O_{22}+O_{32} O_{21}) ) \; ,
\end{eqnarray}
and $g_{Z^\prime Z h}$ can be also found in \cite{Cheung:2010az}.
Thus, the decay rates \eqref{genrate} corresponding to these
processes can be easily obtained from the following amplitudes
squared
\begin{eqnarray}
|\mathcal M_{Z^\prime \to H^+H^-}|^2 &=& \frac{1}{3} g^2_{Z^\prime H^+ H^-} \;  M^2_{Z^\prime} \left( 1- \frac{4 M_{H^{\pm}}^2}{M^2_{Z^\prime}} \right) \nonumber \\
|\mathcal M_{Z^\prime \to HA}|^2 &=& \frac{1}{3} g^2_{Z^\prime H A} \left[ \frac{(M^2_H-M^2_A)^2}{M^2_{Z^\prime}} + M^2_{Z^\prime} \left( 1- \frac{2(M_H^2+M_A^2)}{M^2_{Z^\prime}} \right)  \right] \nonumber \\
|\mathcal M_{Z^\prime \to Zh}|^2 &=& \frac{1}{3} g^2_{Z^\prime Z h} \left[  \frac{(M^2_{Z^\prime} + M_Z^2 - M_h^2)^2}{4M_Z^2M_{Z^\prime}^2}+ 2 \right] \; .
\end{eqnarray}
Finally, for the decay of $Z^\prime$ into $W^{\pm}$ bosons, we have obtained a formula similar to that
of~\cite{Barger:1987xw}, with the coupling in the model discussed here being different by the SM coupling
between $Z$ and $W^{\pm}$ bosons by a factor $\frac{O_{31}}{\cos(\theta_w)}$. Suppression by $O_{31}$ of
the decay $Z^\prime \rightarrow W^{+} W^{-}$, resulting in a branching ratio of a few percent for this mode,
is seen in~\cite{Feldman:2006wb}.

\subsection{Results}

\begin{figure}
\centerline{\epsfig{figure=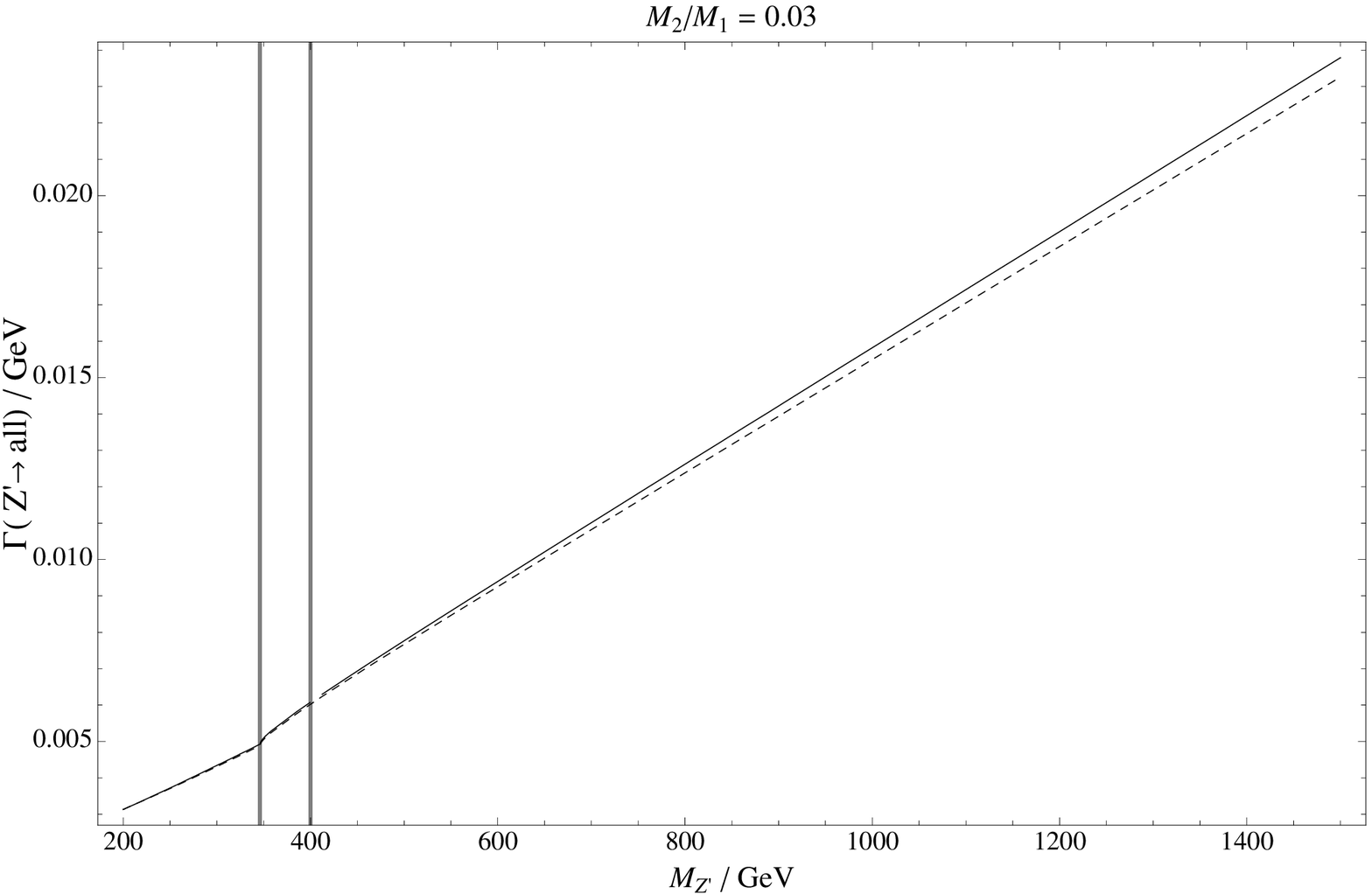,height=8cm,angle=0}}
\caption{Total $\Gamma$ of $Z'$ depending on $M_{Z'}$ for one Higgs doublet (dashed) and two Higgs
doublet model (for Set 1 and $M_2/M_1=0.03$). The vertical lines are kinematic thresholds corresponding to twice the masses of the top quark and charged Higgs of Set 1.\label{totalG003}}
\end{figure}

\begin{figure}
\centerline{\epsfig{figure=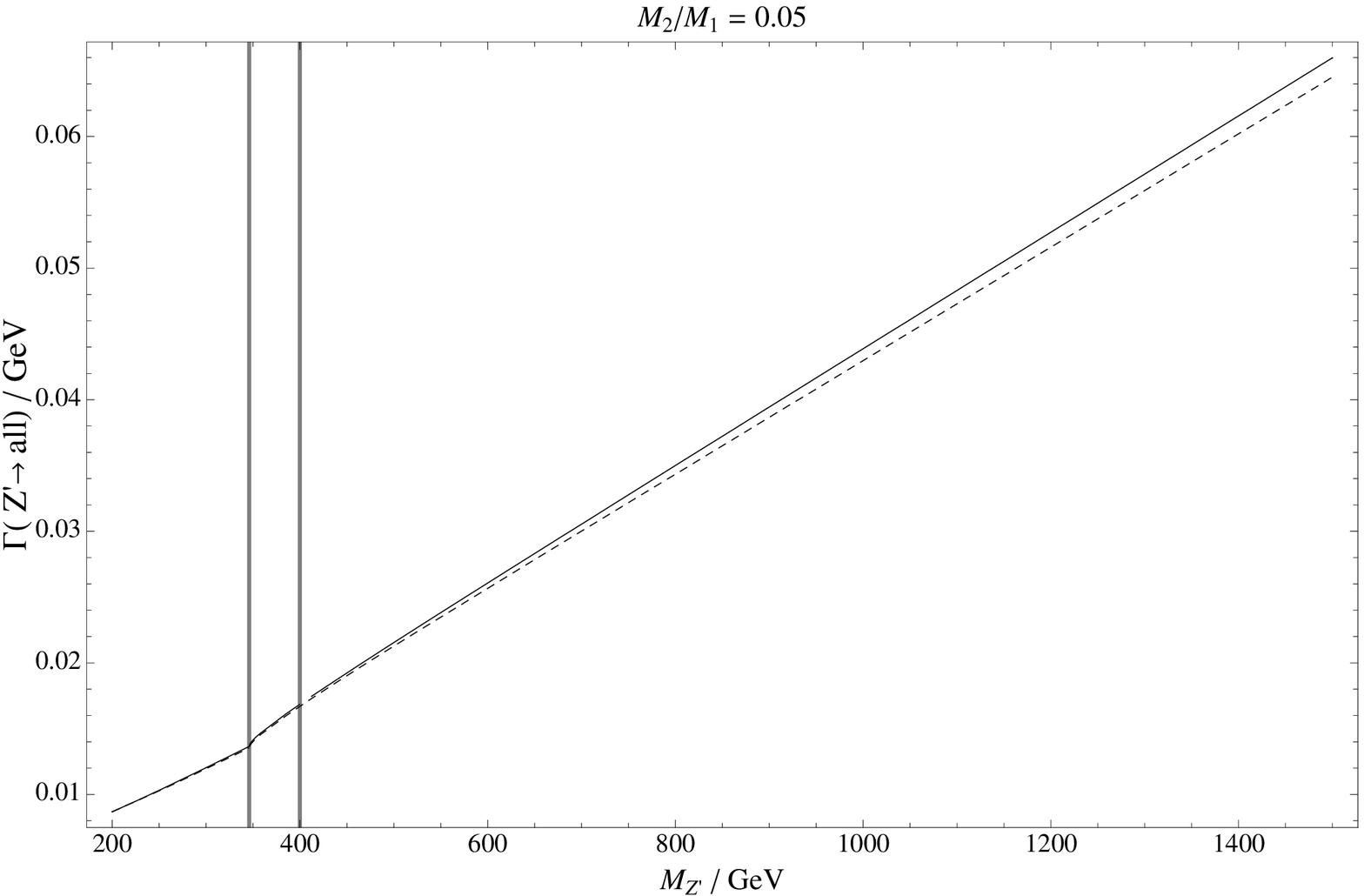,height=8cm,angle=0}}
\caption{Total $\Gamma$ of $Z'$ depending on $M_{Z'}$ for one Higgs doublet (dashed) and two Higgs
doublet model (for Set 1 and $M_2/M_1=0.05$). The vertical lines are kinematic thresholds corresponding to twice the masses of the top quark and charged Higgs of Set 1.\label{totalG005}}
\end{figure}

\begin{figure}
\centerline{\epsfig{figure=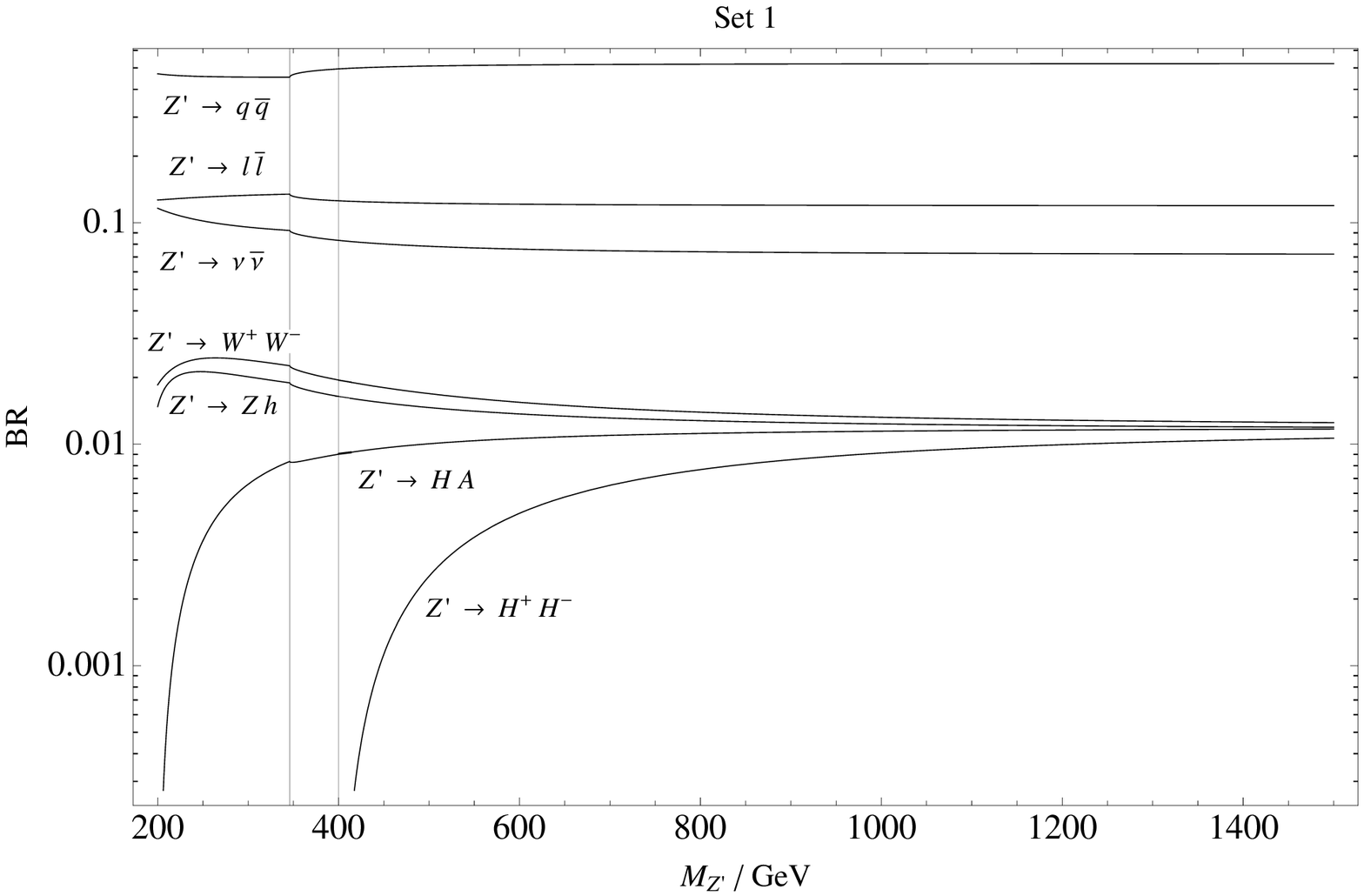,height=8cm,angle=0}}
\caption{Branching ratios depending on $M_{Z'}$ for Set 1. All the decay channels into quarks have been
considered together as a single quark channel. The vertical lines are kinematic thresholds corresponding to twice the masses of the top quark and charged Higgs of Set 1.\label{allset1}}
\end{figure}

\begin{figure}
\centerline{\epsfig{figure=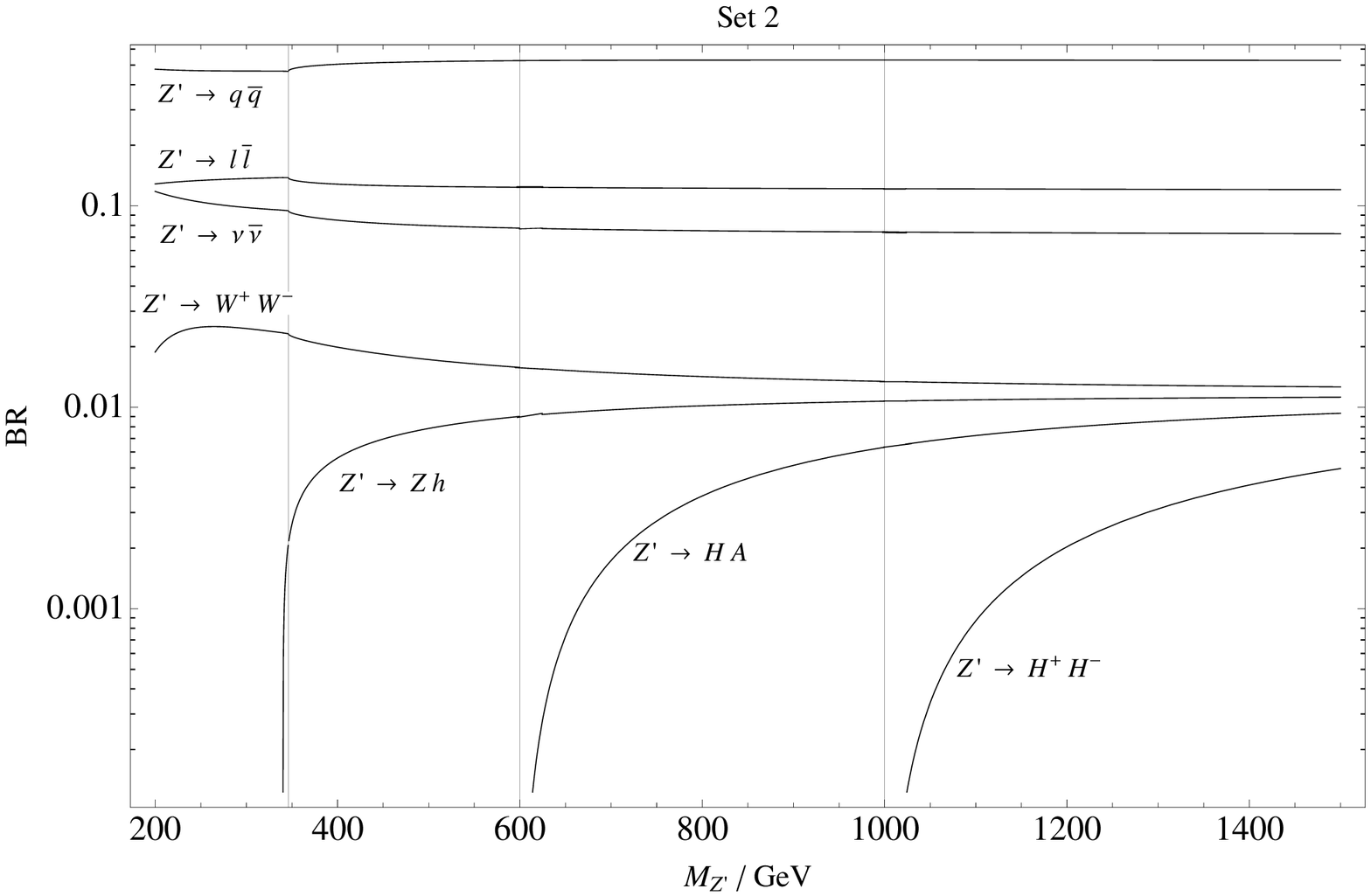,height=8cm,angle=0}}
\caption{Branching ratios depending on $M_{Z'}$ for Set 2. All the decay channels into quarks have been
considered together as a single quark channel. The vertical lines are kinematic thresholds corresponding to twice the masses of the top quark, the neutral Higgs and charged Higgs of Set 2.\label{allset2}}
\end{figure}

Our results are summarized in the figures below. We have fixed the Higgs boson masses (Set 1 and
Set 2 as can be seen in Table~2), as well as the coupling constant $g_X$ considering two cases,
one in which the coupling is small, $g_X=0.001$, and one in which the coupling is comparable to the SM couplings, $g_X=0.1$. Then the only free parameter left is the heavy gauge boson mass. Therefore, in the figures shown below the
independent variable is the mass of $Z^\prime$. First we focus on the case where $g_X=0.001$.
Figures \ref{totalG003} and \ref{totalG005} show the total decay width of $Z^\prime$ (in GeV)
as a function of its mass for Set 1, with $M_2/M_1=$ 0.03 and 0.05, respectively. In the rest of the figures the impact of changing the value for the
ratio $M_2/M_1$ is negligible, so it is fixed at 0.03. Figures \ref{allset1} and \ref{allset2} show all branching ratios as a
function of $M_{Z^\prime}$ (for Set 1 and Set 2 respectively). All the decay channels into quarks have been
considered together as a single quark channel. However, we have checked that $Z^\prime$ decay into quarks
is dominated by the up quark contributions, as in Figure~1 of~\cite{Feldman:2006wb}. The straight vertical
lines correspond to the thresholds, one for the top quark ($\sim 346$ GeV), one for the neutral Higgs bosons
(600 GeV for Set 2 only) and one for the charged Higgs bosons (400 GeV for Set 1 and 1000 GeV for Set 2). We
remind the reader that in the SM, the branching ratio of the $Z$ boson to electrons or muons
or tau leptons is 0.034 for each of them, to all neutrino species (invisible channel) is 0.2, and to
hadrons is 0.7. To compare with the Stueckelberg extension of the SM with just one Higgs doublet,
we show for that model the branching ratios of $Z^\prime$ in Figure \ref{all1H} (for Set 1), and the total
decay width in Figures \ref{totalG003} and \ref{totalG005} together with the total width for our model with two Higgs doublets.
In the model with one Higgs doublet there are no decay channels to inert Higgs bosons, and for a large
enough $M_{Z^\prime}$, where the branching ratios of $Z^\prime$ to the inert Higgs bosons become
significant, the decay widths in the two models tend to differ. However, the difference is small since
the dominant contribution to the decay width is from $Z^\prime$ to fermions, which scales as
$M_{Z^\prime}g_Y^2(M_2/M_1)^2$~\cite{Kors:2005uz}. Furthermore, in the model with
one Higgs doublet only, there is just the SM neutral Higgs boson, while in the model with
two Higgs doublets there are both neutral and charged Higgs bosons.
\begin{table}[h!]
\begin{center}
\begin{tabular}{|c|c|c|}
\hline
&Set 1& Set 2 \\
\hline
$M_{H^{\pm}}$ (GeV) &200& 500\\
$M_{H,A}$ (GeV) &100& 300\\
$M_{h}$ (GeV) &100& 250\\
\hline
\end{tabular}
\caption{The two sets of Higgs boson masses used in the analysis.}
\end{center}
\end{table}
Clearly, if a charged Higgs
boson is seen at colliders, this would be a direct evidence of physics beyond the SM.
Without Yukawa couplings the
charged Higgs bosons cannot directly decay into fermions, and therefore
the dominant decay channels of the charged Higgs bosons are just two, $H^{\pm} \rightarrow W^{\pm} \: H$
and $H^{\pm} \rightarrow W^{\pm} \: A$. Taking into account that $H$ and $A$ are degenerate in mass,
the model discussed here predicts that there are two main decay channels for $H^{\pm}$ with the two branching
ratios being equal to $1/2$. A detailed discussion of the Higgs phenomenology is postponed to a future work \cite{wip}. We can also mention here in passing that if the decay channel $h \rightarrow ZZ$ is kinematically
allowed, the SM Higgs boson can be easily found through the so-called four-lepton golden Higgs channel,
$h \rightarrow ZZ \rightarrow l^+ l^- l^+ l^-$~\cite{guide}.

As in the case with one
Higgs doublet, the total decay width is much smaller than in other models~\cite{Anoka:2004vf,Aydemir:2009zz},
and therefore a heavy gauge boson is expected to show up at colliders as a sharp resonance. Finally,
in Figure \ref{BrBr} we show ratios of decay widths of two channels as a function of $M_{Z^\prime}$, and
in particular we have chosen to show the
following ratios: Leptons to hadrons, leptons to neutrinos, charged Higgs to neutral Higgs, and
$W^{\pm}$ bosons to SM Higgs and $Z$ boson. Recall that in the SM the ratio of leptons to neutrinos is
0.17, and the ratio of leptons to hadrons is 0.05.

Finally, notice that in Figures \ref{totalG003} and \ref{totalG005}, although they look very
similar, the scale is different. When the ratio $M_2/M_1$ is increased from 0.03 to
0.05, the total decay width also increases by a factor $\sim 3$, because the couplings of the new
gauge boson are now larger. We have also checked that the plot with larger mass ratio showing the branching fractions
cannot be distinguished from the one with smaller mass ratio.

\begin{figure}
\centerline{\epsfig{figure=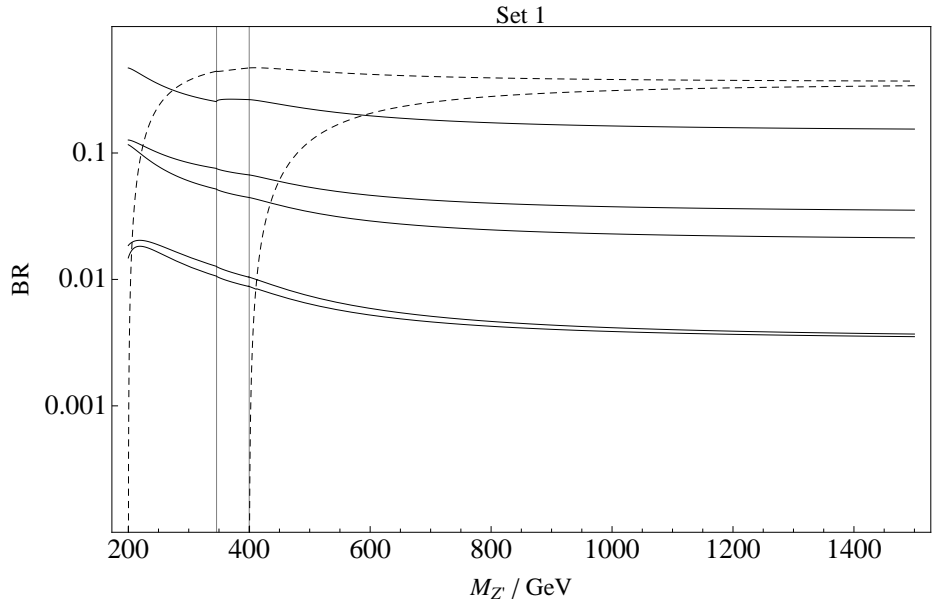,height=8cm,angle=0}}
\caption{Same as Figure \ref{allset1} but only changing the value of $Y_Xg_X$ to $+0.1$. Apart from the total decay rate, which is larger, the only relevant difference is for the branching ratios of $Z'$ decaying into inert Higgs (dashed lines), that increase significantly.\label{allset1large}}
\end{figure}

\begin{figure}
\centerline{\epsfig{figure=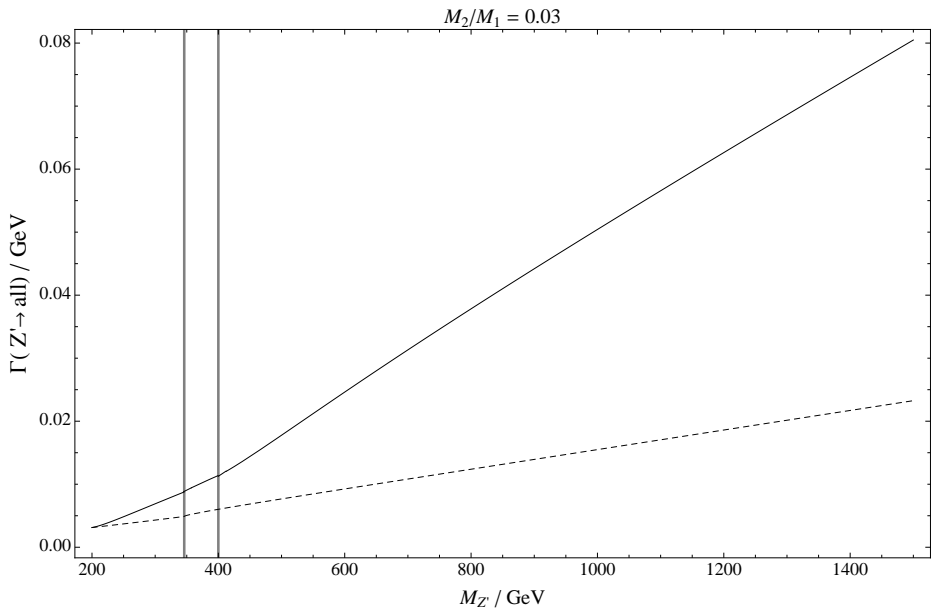,height=8cm,angle=0}}
\caption{Same as Figure \ref{totalG003} but only changing the value of $Y_Xg_X$ to $+0.1$. The total decay rate when the inert Higgs are present (two-Higgs-doublet model) increases.\label{totalG003large}}
\end{figure}

\begin{figure}
\centerline{\epsfig{figure=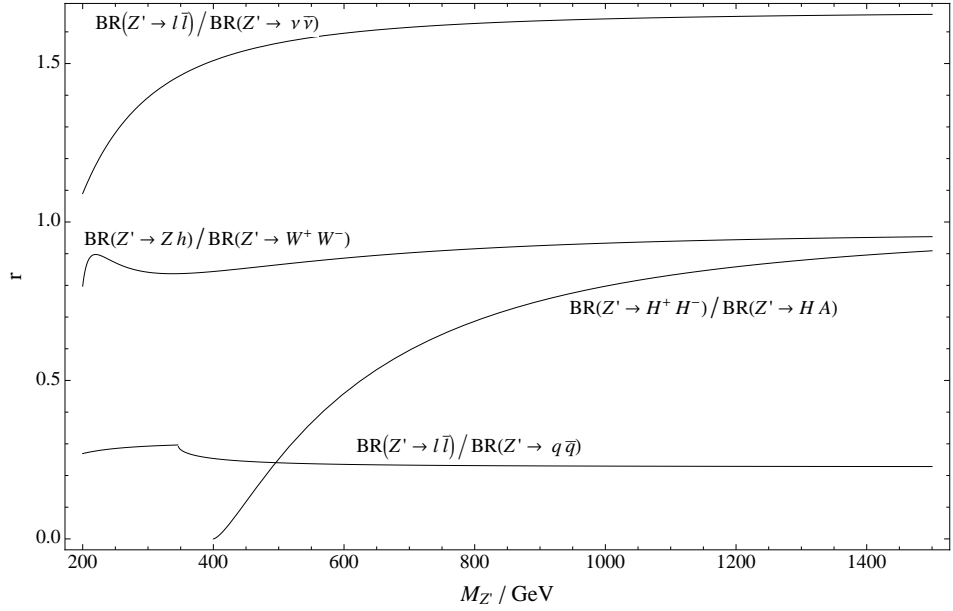,height=8cm,angle=0}}
\caption{Ratios $r$ of partial decay widths depending on $M_{Z'}$ (for Set 1).\label{BrBr}}
\end{figure}

\begin{figure}
\centerline{\epsfig{figure=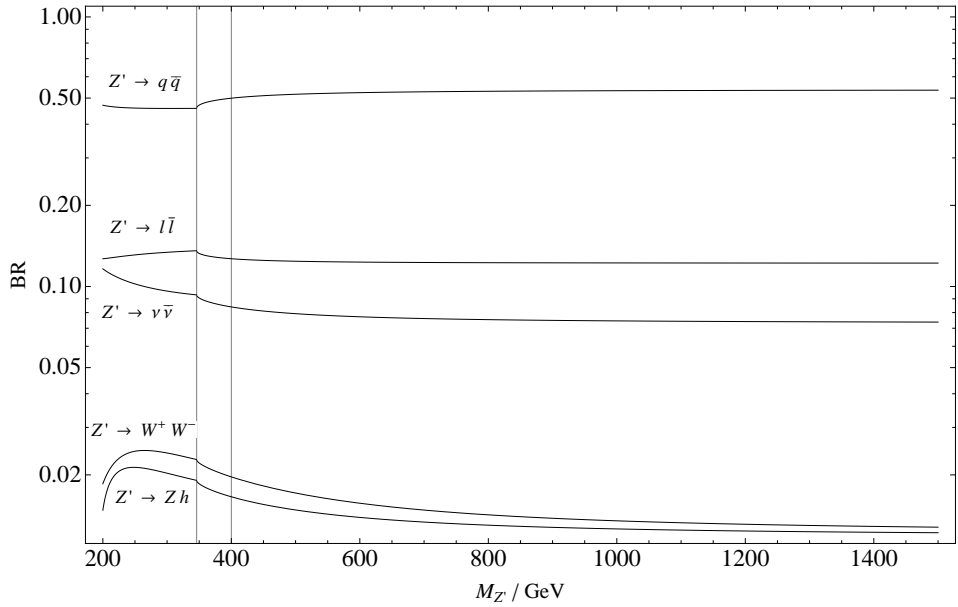,height=8cm,angle=0}}
\caption{Branching ratios depending on $M_{Z'}$ in the case of one Higgs doublet model (for Set 1). The vertical lines are kinematic thresholds corresponding to twice the masses of the top quark and charged Higgs of Set 1.\label{all1H}}
\end{figure}

We now consider the case where $g_X=0.1$ for Set 1 and $M_2/M_1=0.03$. Most of the decay modes remain the same, apart from the ones into the inert Higgs bosons, for which the coupling now is larger, leading to
larger partial decay widths. Figure \ref{totalG003large}
shows the effect of increasing the coupling constant $g_X$ in the total decay width, while Figure
 \ref{allset1large} shows the effect on the branching ratios. The curves corresponding to the decays
into the inert Higgs bosons preserve their shape, but now they are above the rest. The sign of $Y_X$
has been taken to be positive. If we change the sign of $Y_X$ we obtain a similar plot where
the branching ratios for the inert Higgs bosons are slightly larger.

\section{Conclusions}

A model with an extra $U(1)$ and a second Higgs doublet has been investigated.
It is assumed that the fermions and the SM Higgs are neutral under the extra
$U(1)$, while the dark Higgs is charged. Thus, Yukawa couplings for the additional
Higgs are not allowed, and the FCNC problem is avoided. From this point
of view the model is similar to the inert 2HDM, although the gauge symmetry
is more restrictive than the $\mathcal Z_2$ discrete symmetry. The massive gauge bosons
obtain their masses from two separate mechanisms, namely from the usual Higgs mechanism,
as well as from the Stueckelberg mechanism. The interplay between
the heavy gauge boson and the extended Higgs sector makes the phenomenology
of this model very rich. We have computed the total decay width and all the branching
ratios of $Z'$ as a function of its mass for two different sets of the Higgs bosons masses.
We find that two distinct features of the model are a) a sharp decay width for the heavy
gauge boson, characteristic of the Stueckelberg mechanism like in the corresponding model
with just one Higgs doublet, and b) a pair of charged Higgs bosons with no Yukawa couplings
decaying dominantly into a $W^{\pm}$ boson and a neutral Higgs boson $H$ or $A$, with the two branching
ratios being equal to $1/2$ each.

\section*{Acknowledgments}

We wish to thank L.~L.~Honorez and M.~Gustafsson for correspondence, and also A.~Pich for
reading the manuscript and useful comments. G.~P.~acknowledges
financial support from FPA2008-02878, and
Generalitat Valenciana under the grant PROMETEO/2008/004.
The work of P.~T.~has
been supported in part by MEC (Spain) [FPU AP2006-04522 and Grant FPA2007-60323] and by the Spanish
Consolider Ingenio 2010 Programme CPAN [CSD2007-00042].

\bibliography{bibliography}

\newpage

\end{document}